\documentclass[global,twocolumn]{svjour}
\usepackage{graphicx}
\journalname{Applied Physics B}
\usepackage{times}

\makeatletter
\def\lambdabar{\protect\@lambdabar}
\def\@lambdabar{%
\relax
\bgroup
\def\@tempa{\hbox{\raise.73\ht0
\hbox to0pt{\kern.25\wd0\vrule width.5\wd0
height.1pt depth.1pt\hss}\box0}}%
\mathchoice{\setbox0\hbox{$\displaystyle\lambda$}\@tempa}%
{\setbox0\hbox{$\textstyle\lambda$}\@tempa}%
{\setbox0\hbox{$\scriptstyle\lambda$}\@tempa}%
{\setbox0\hbox{$\scriptscriptstyle\lambda$}\@tempa}%
\egroup
}
\makeatother

\begin{document}

\title{Fundamental Limits for Coherent Manipulation on Atom Chips}

\author{Carsten Henkel$^1$
\and Peter Kr\"uger$^2$
\and Ron Folman$^2$
\and J\"org Schmiedmayer$^2$
}

\institute{$^1$Institut f\"ur Physik,
Universit\"at Potsdam,
14469 Potsdam, Germany,
\email{Carsten.Henkel@quantum.physik.uni-potsdam.de}
%\\
\\
$^2$Physikalisches Institut,
Universit\"at Heidelberg,
69120 Heidelberg, Germany
%\\
}

\date{22 August 2002, revised 30 October 2002, accepted 11 December 2002}

\maketitle

\begin{abstract}
%\abstract{
The limitations for the coherent manipulation of neutral atoms
with fabricated solid state devices, so-called `atom chips', are
addressed. Specifically, we examine the dominant decoherence
mechanism, which is due to the magnetic noise originating from the
surface of the atom chip. It is shown that the contribution of 
fluctuations in the chip wires at the shot noise level is not
negligible. We estimate the coherence times and
discuss ways to increase them. Our main conclusion is that future
advances should allow for coherence times as long as $1$ second,
a few $\mu$m away from the surface.
\\[1mm]
%}
\end{abstract}

\titlerunning{Limits for Atom Chips}
\authorrunning{C. Henkel et al.}

%\maketitle

%%% introduction for Osnabrueck 2002 proceedings

%\section*{Introduction}

\noindent
In the quest for physical implementations of quantum information
processing, ``atom chips'' are currently of great interest. This
is because they promise well-controlled quantum optical manipulations
of neutral atoms in integrated and scalable microtrap arrays.
In these traps, atoms are strongly confined by
electromagnetic fields close to nanostructured solid-state
substrates. Microtraps used in current experiments are 
magnetic traps produced by current-carrying wires
\cite{Haensch98,Reichel99,Zimmermann00a,Reichel01c,%
Reichel01b,Zimmermann01,Ketterle02a,Schmiedmayer02b,%
Reichel02a,Folman02}
and periodically magnetized substrates \cite{Hinds99b},
or hybrid traps involving optical or electric fields
\cite{Schmiedmayer98a,Grimm00b}.

In this paper, we discuss the limitations that wire-based
magnetic traps on atom chips may have to face when they
are downscaled into the micron range.
Recently, both theoretical and experimental
indications have been found that the `hot' chip substrate
--- typically held at room temperature ---
is not a quiet environment: at distances below
a few 100$\,\mu$m from the chip, the trap lifetime
is shorter than in free space and the atom temperature increases
\cite{Reichel01b,Zimmermann01,Henkel99b,Henkel99c,Zimmermann02a}.
While it is not excluded that strong compression
in these microtraps plays a role
due to enhanced collisional interactions
(see \cite{Weiner99} for a review),
noisy magnetic fields may also be involved.
They provide a coupling to the environment that may cause
loss, heating and decoherence and are elaborated upon
in this paper. We review the sources of magnetic fields and
quote estimates for trapping and coherence times.
In particular, we discuss the contribution
of electric current noise at the shot noise level
and evaluate its spectral and spatial properties.
This is compared to the noise due to the thermal chip substrate.

\section{Atom chip `building block': the side guide}

\begin{figure}[h]
\centerline{%
\resizebox{80mm}{!}{%
\includegraphics*{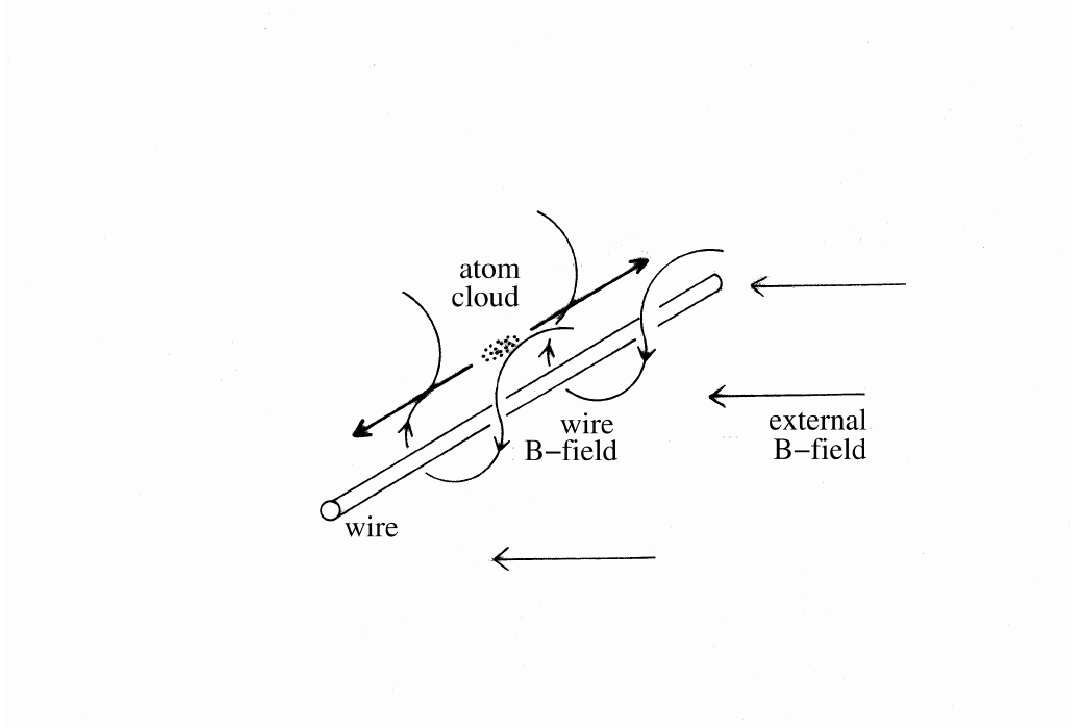}}}
\caption[]{Principle of a linear magnetic quadrupole guide
(``side guide"). Figure courtesy of Sierk P\"otting.}
\label{fig:guide.illu.eps}
\end{figure}

In the 1930's, Frisch and Segr\'e realized that when a
homogeneous magnetic field (`bias field') is superimposed with
the field of a straight wire current,
the magnetic field vanishes on a line parallel to the current
(see figure~\ref{fig:guide.illu.eps})
\cite{Frisch33}. In the vicinity of this line,
the field increases in a quadrupolar fashion.
The height of the field zero is given by
\begin{equation}
h = \frac{ \mu_0 }{ 2 \pi }
\frac{ I }{ B_b },
\label{eq:guide-height}
\end{equation}
where $\mu_0 = 4\pi\,{\rm mm}\,{\rm G} / {\rm A}$
is the vacuum permeability, $I$ the wire current
and $B_b$ the bias field magnitude. (This expression applies to
finite size wires provided their diameter $\ll h$.) This `side
guide' can be implemented on an atom chip using a lithographically
etched wire on the chip surface. The setup is ideal for miniaturization
since the guide height is reduced using smaller currents
(with less ohmic dissipation).

Atoms with a nonzero magnetic moment
{\boldmath $\mu$} are trapped around the magnetic field minimum
provided {\boldmath $\mu$} keeps an antiparallel orientation
with respect to the local magnetic field (adiabatic approximation).
These `weak field seekers' are attracted to the field
zero and get trapped in a potential,
$\mu_{\Vert} |{\bf B}_0( {\bf x} ) |$, proportional to
the magnitude of the magnetic field.
(The subscript 0 distinguishes
the static trapping field from the field fluctuations discussed
later.)
It is particularly
interesting to introduce a longitudinal magnetic field (along the wire)
so that the field magnitude never reaches zero: this helps satisfying
the adiabatic approximation everywhere and reduces so-called `Majorana
flips' into untrapped magnetic sublevels.
The trapping potential is then a harmonic well instead of
a linear one around the guide center. This is the guide geometry
we focus on in the following. The relevant parameters are the
trap height $h$ (Eq.\ref{eq:guide-height}), the oscillation
frequency $\Omega/2\pi$ in the harmonic well, and the Larmor
frequency $\omega_{\rm L} = \mu_\Vert |{\bf B}_0( {\bf r})| / \hbar$
where ${\bf B}_0( {\bf r})$ is the field at the trap center
and $\mu_\Vert$ is the magnetic moment along the center field
\cite{Reichel01c,Folman02}. For the lowest quantum states in
the potential well, the magnetic field is predominantly
longitudinal, and the single-particle wave functions are
approximately harmonic oscillator states for quantum numbers
up to $n \sim \omega_{\rm L} / \Omega$.

The validity of the adiabatic approximation is also determined 
by the ratio $\omega_{\rm L} / \Omega$: it has been estimated that
nonadiabatic losses are exponentially suppressed when this
ratio is large \cite{Sukumar97} (see also \cite{Eberlein00a}).
Similarly, tunnelling losses down to the chip surface can
be made exponentially small with a sufficiently high
and `thick' potential barrier. The main loss channel is then
provided by the uncontrollable coupling to the surface via
magnetic noise.

\section{Interaction with magnetic noise}

\subsection{Noise spectrum and spin flip rate}

The coupling of the atomic magnetic moment to fluctuating magnetic
fields gives rise to both spin flips and changes in the center-of-mass
motion (scattering). The rate of these processes is given by
Fermi's Golden Rule. We recall here that it can be conveniently
expressed in terms of the noise spectrum of the magnetic field
fluctuations.
(See \cite{Sipe84} for a similar approach and Chap.\ IV
of \cite{CDG2} for the derivation of a full master equation.)

If we write $|{\rm i} \rangle$ and $|{\rm f}\rangle$ for the atomic
states before and after the transition, the transition rate is
\begin{equation}
\Gamma_{\rm i \to f} =
\frac{ 2 \pi }{ \hbar }
\sum_{\rm F,I}
p( {\rm I} )
\left|
\langle {\rm F, f} |
H_{\rm int}
| {\rm I, i} \rangle
\right|^2
\delta(
E_{\rm F}
+
E_{\rm f}
-
E_{\rm I}
-
E_{\rm i}
)
,
\label{eq:Gamma-if-Fermi}
\end{equation}
where $|{\rm I} \rangle$ and $|{\rm F}\rangle$ are initial and final
states for the field, the summation being an average over the initial
field states (with probabilities $p({\rm I})$) and a trace over the final
field states.  The interaction Hamiltonian is given by
$H_{\rm int} = - \mbox{\boldmath$\mu$}\cdot{\bf B}( {\bf x} )$.

Consider first the rate for spin flips. Since only a
subset of magnetic sublevels $|m_{\rm i}\rangle$ are weak field
seekers, spin flips $|m_{\rm i}\rangle \to |m_{\rm f}\rangle$
are responsible for trap loss. The magnetic field is
evaluated at the position ${\bf r}$ of the trap center. (An
average over the atomic position distribution would be more
accurate.)
We write the $\delta$-function for energy conservation as a time
integral
over ${\rm e}^{{\rm i}(
E_{\rm I}
-
E_{\rm F}
-
\hbar \omega_{\rm f\-i})
t / \hbar}$
where
$\hbar\omega_{\rm f\-i} = E_{\rm f} - E_{\rm i}$.
The exponential ${\rm e}^{ {\rm i} ( E_{\rm I} - E_{\rm F})
t / \hbar}$
can be removed by introducing
the field operators in the Heisenberg picture
\begin{equation}
{\bf B}( {\bf r}, t ) =
{\rm e}^{ {\rm i} H_0 t / \hbar }
{\bf B}( {\bf r} )
{\rm e}^{ -{\rm i} H_0 t / \hbar }
\end{equation}
(here, $H_0$ is the free field Hamiltonian)
and taking matrix elements of this operator between the initial
and final field states. This gives
\begin{equation}
{\rm e}^{ {\rm i} ( E_{\rm I} - E_{\rm F}) t / \hbar }
\langle {\rm I} |
{\bf B}( {\bf r} )
| {\rm F} \rangle
=
\langle {\rm I} |
{\bf B}( {\bf r}, t )
| {\rm F} \rangle
.
\end{equation}
The sum over the final states $| {\rm F}\rangle$ now reduces
to a completeness relation and we get ($\alpha,\beta$ denote
field components)
\begin{eqnarray}
&&
2\pi\hbar
\sum_{\rm F,I}
p( {\rm I} )
\langle {\rm I} |
{B}_\alpha( {\bf r} )
| {\rm F} \rangle
\langle {\rm F} |
{B}_\beta( {\bf r} )
| {\rm I} \rangle
\delta(
E_{\rm F}
-
E_{\rm I}
- \hbar \omega
)
\nonumber
\\
&& =
\int\limits_{-\infty}^{\infty}
\!{\rm d}t \,
{\rm e}^{ {\rm i}\omega t }
\sum_{\rm I}
p( {\rm I} )
\langle {\rm I} |
{B}_\alpha( {\bf r}, t )
{B}_\beta( {\bf r}, 0 )
| {\rm I} \rangle
\nonumber
\\
&& =
S_{\alpha\beta}( {\bf r}; \omega )
.
\label{eq:def-noise-spectrum}
\end{eqnarray}
In the last line, we have defined the magnetic noise spectrum
which is the Fourier transform of the
field's autocorrelation function.
The rate for spin flips can now be written as
\begin{equation}
\Gamma_{\rm i \to f} =
\frac{ 1 }{ \hbar^2 }
\sum_{\alpha, \beta = x,y,z}
\langle m_{\rm i} | \mu_\alpha | m_{\rm f} \rangle
\langle m_{\rm f} | \mu_\beta | m_{\rm i} \rangle
S_{\alpha\beta}( {\bf r}, -\omega_{\rm f\-i} )
.
\label{eq:rate-spectrum}
\end{equation}
Since $m_{\rm f} \ne m_{\rm i}$, the matrix elements of
$\mu_{\alpha}$ are only nonzero for directions perpendicular to the
magnetic field at the trap center. We also recover the selection rule
$m_{\rm f} - m_{\rm i} = \pm 1$ so that the relevant transition
frequency is the Larmor frequency $|\omega_{\rm f\-i}| = \omega_{\rm L}$.
The spin flip rate gives the order of
magnitude of trap loss even if more than one weak-field seeking
Zeeman states, $m_{\rm i} = +2, +1$, say, are trapped
(possible with many of the alkali atoms).
This is because the
matrix elements between adjacent sublevels do not significantly differ
in magnitude so that the atoms reach the non-trapped sublevel
$m_{\rm f} = 0$ after a time $ \sim 2/\Gamma_{+2 \to +1}$.

We finally note that as long as the behaviour of the
`environment' (the field) is ignored in the description
of the atom's dynamics, the noise spectrum is the only quantity
needed to characterize the environment.
It is also an experimentally
measurable quantity: for example, the rms magnetic noise
$\langle B_{x}^2( {\bf r} ) \rangle^{1/2}$
measured by a spectrum analyzer in a given frequency band
$\Delta \omega / 2\pi$ around $\omega$
is $( 2 S_{xx}( {\bf r}, \omega ) \, \Delta \omega / 2\pi )^{1/2}$,
the factor $2$ accounting for the sum over positive and negative
frequencies. The atomic spin flip rate may be regarded as
an alternative way to measure the noise spectrum. In order
of magnitude,
the magnetic moment is comparable to the Bohr magneton,
$\mu_{\rm B}$ 
($\mu_{\rm B} / 2\pi\hbar = 1.4\,$MHz/G), and we get
\begin{equation}
\Gamma_{\rm i\to f}( {\bf r} ) \sim
0.01\,{\rm s}^{-1}
( \mu / \mu_{\rm B} )^2
\frac{ S_{\alpha\beta}( {\bf r}, \omega_{\rm L} ) } {
{\rm pT}^2 / {\rm Hz} }
.
\end{equation}
Note that current SQUID magnetometers are able to detect magnetic field
noise even on the $10\,{\rm fT}/\sqrt{ {\rm Hz} }$ scale
\cite{Varpula84}.

\subsection{Near field noise}

\begin{figure}[h]
\centerline{%
\resizebox{80mm}{!}{%
\includegraphics*{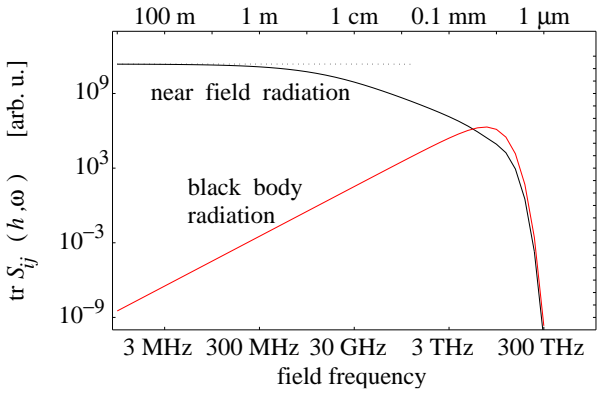}}}
\caption[]{Magnetic near field noise spectrum. The near field
spectrum is calculated along the lines given in \cite{Henkel99c}.
The source is a copper half-space at $T = 300\,$K, observation
distance $h = 1\,\mu$m. The trace of the magnetic correlation
tensor is shown. The top labels give the
wavelength $\lambda$.}
\label{fig:magn.spec.nf.eps}
\end{figure}

The magnetic noise spectrum close to a solid substrate shows
dramatic differences with respect to the well-known blackbody
spectrum. This is because thermally excited currents in the
substrate (Johnson-Nyquist noise) generate electromagnetic
fields with substantial nonpropagating components.
As an illustration, figure~\ref{fig:magn.spec.nf.eps}
shows the noise spectrum at a fixed distance ($h = 1\,\mu$m)
above a metallic half-space, calculated along the lines of
\cite{Henkel99c}.
One gets an increase of several orders of
magnitude for low frequencies (wavelength $\lambda$ much
larger than $h$). In addition, the spectrum is rather flat in this
range. Only at high frequencies ($\lambda < h$), the Planck
spectrum is recovered because the detector then enters the
far field of the source.

\begin{figure}[h]
\centerline{%
\resizebox{80mm}{!}{%
\includegraphics*{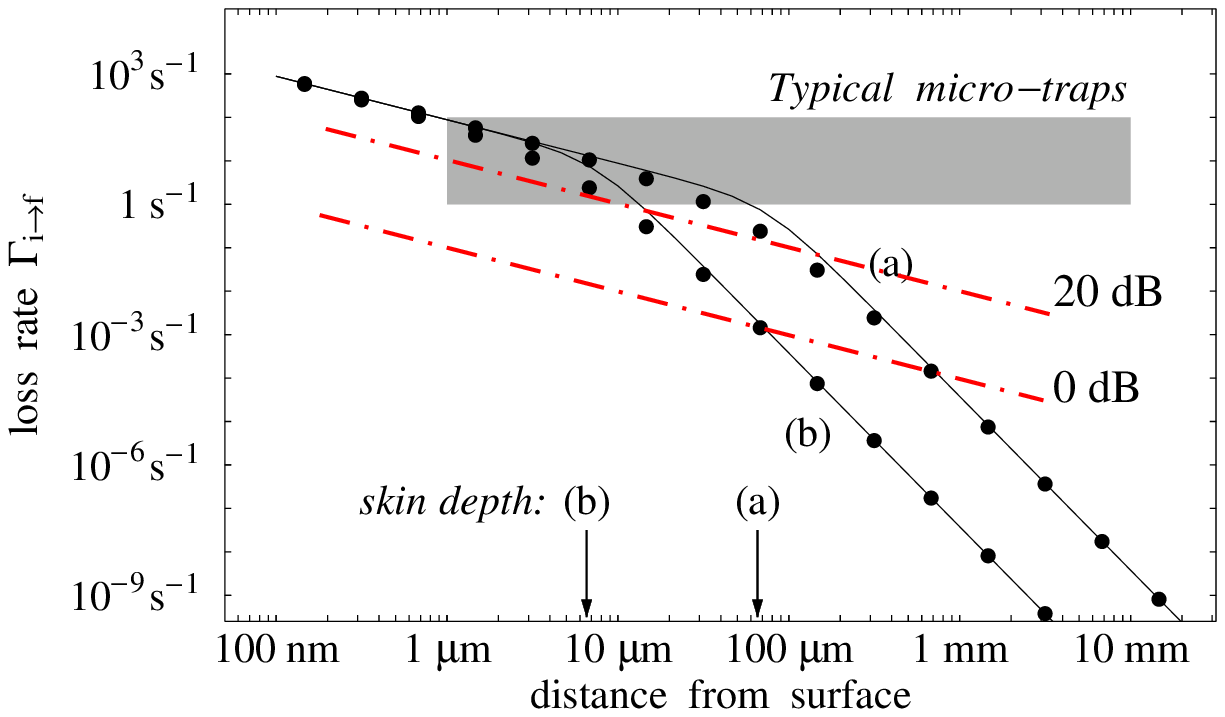}}}
\caption[]{Rate of spin flips in a microtrap above a planar
metal substrate. The black solid lines describe the near
field noise
generated by thermally excited currents (Johnson noise) in
the substrate, as given by Eq.(\ref{eq:nf-noise-spectrum}).
The dots are an exact numerical calculation from~\cite{Henkel99c}.
Larmor frequencies
$\omega_{L} = \mu_{B} |{\bf B}( {\bf r} )| / \hbar =
1\,{\rm MHz}\times 2\pi$ (a) and $100\,{\rm MHz}\times 2\pi$ (b)
are chosen.
The red dash-dotted lines describe the noise due to fluctuations
in the electric current of a side guide. The current noise
is assumed at shot noise level (0 dB) or 20 dB above shot noise.
The guide height $h$ is lowered by ramping down the wire current $I$
with a constant ratio $I/h$ and
at fixed bias field $B_b = 100\,$G,
\emph{cf.}\ Eq.(\ref{eq:guide-height}).}
\label{fig:flip.rate.eps}
\end{figure}

In the low-frequency range, the magnetic near field noise above a
planar substrate is approximately characterized by the spectrum
\cite{Henkel99c}
\begin{equation}
    \frac{ \omega h }{ c } \ll 1: \quad
    S_{\alpha\beta}( {\bf r}, \omega ) =
    \frac{ \mu_{0}^2 k_{B} T }{ 16 \pi \,\varrho }
    \frac{ s_{\alpha\beta} }{ h }
    \left[ 1 + \frac{ 2 \,h^3 }{ 3 \,\delta(|\omega|)^3 }
    \right]^{-1}
    ,
\label{eq:nf-noise-spectrum}
\end{equation}
where $T, \varrho$ are the substrate temperature and resistivity,
respectively, $h$ is the observation distance,
$s_{\alpha\beta} = {\rm diag}( \frac12, 1, \frac12 )$ is a diagonal
tensor (the distinguished axis is along the surface normal), and
$\delta( \omega ) =
\sqrt{ 2 \varrho / \mu_0 \omega }$ is the skin depth.
In order of magnitude, the corresponding
spin flip rate is surprisingly large for traps at a micrometer
distance:
\begin{equation}
h \ll \delta( \omega_{\rm L} ): \quad
\Gamma_{\rm i \to f} \sim 100\,{\rm s}^{-1}
\frac{ ( \mu / \mu_{\rm B} )^2 (T / 300\,{\rm K})
}{
( \varrho / \varrho_{\rm Cu} ) ( h / \mu{\rm m} ) }
\label{eq:estimate-gamma}
,
\end{equation}
where $\varrho_{\rm Cu} = 1.7 \times 10^{-6}{\rm\Omega\,cm}$ is
the copper resistivity. The flip rate is plotted as a function of
distance in figure~\ref{fig:flip.rate.eps}. A larger Larmor
frequency (longitudinal bias field) only helps reducing spin
flips when $h > \delta( \omega_{\rm L} )$, giving a scaling
$\Gamma_{\rm i\to f} \propto \omega_{\rm L}^{-3/2}$. 
Finally, 
there are many plausible reasons to expect that the 
linear dependence on~$T$ of the flip rate~(\ref{eq:estimate-gamma})
does not continue down to very low temperatures: 
the full Bose-Einstein occupation number 
has to be used, $ k_{\rm B} T \mapsto 
\hbar\omega_{\rm L}\, (1 - {\rm e}^{ -\hbar \omega_{\rm L} / k_{\rm B} T }
)^{-1}$, and other sources of magnetic fields may come into
play like spin waves, lattice vibrations etc.
One thus probably gets lifetimes (and coherence times,
see the following) shorter than predicted 
by~(\ref{eq:estimate-gamma}).

At room temperature,
the estimate~(\ref{eq:estimate-gamma}) describes a worst case
because most materials have a resistivity larger than copper.
A structured substrate like a thin metallic layer or a wire
also helps: the magnetic noise then
decreases more rapidly with distance $h$ \cite{Folman02,Henkel01a}.
As a general rule, the smaller the amount of metallic material, the
lower the magnetic noise. This can be understood from a simple model
where thermally excited currents in each volume
element of the conducting
substrate contribute to the total magnetic noise above
(a similar approach has been used in \cite{Varpula84,Turchette00a}).
For distances $h \ll \delta( \omega )$, one obtains
a noise spectrum accurate within a factor of two from an incoherent
addition of magnetostatic fields, neglecting the influence of the
material on the field propagation. For example, at a distance $h$ 
from a single, thin wire with radius $a \ll h$, the theory
of Ref.\cite{Henkel01a} yields a magnetic 
noise tensor given by Eq.(\ref{eq:nf-noise-spectrum}) with the replacement
$s_{\alpha\beta}/h \mapsto (\pi a^2 / h^3) {\rm diag}( 2, \frac12,
\frac32 )$, showing a similar weak anisotropy. The components specify 
the azimuthal, radial and longitudinal directions in cylindrical coordinates,
respectively. 
The correction involving the skin depth $\delta( \omega )$ cannot be 
obtained in the magnetostatic approximation.

Experimental data \cite{Reichel01b,Zimmermann01,Zimmermann02a} indicate
that trap lifetimes as long as or even longer than estimated
by~(\ref{eq:estimate-gamma}) (several $100\,$ms to
$100\,$s) are achievable: these traps were built close to
thin wires ($3$ to $90\,\mu$m width) or
semiconductor substrates covered with thin metallic layers
(thickness $\ll h$), at distances between $20\,\mu$m and $2\,$mm.
It should be noted that
the distance dependence of the trap lifetime measured in
\cite{Zimmermann02a} only agrees qualitatively with our theory,
and more detailed investigations are required.

\subsection{Current noise}

The electric currents that generate the side guide fields are also subject
to fluctuations that drive spin flips and deform the trapping
fields. The impact of technical noise can be reduced using
electronic filtering, ``quiet'' drivers, by correlating the currents
for the wire and the bias field etc.
This works down to the most fundamental level where the noise due
to the discrete value of the electron charge comes into play (`shot noise').
For a current $I$ in free space, shot noise has a frequency-independent
spectrum given by
\begin{equation}
    SN_{I} = e I \approx 0.16 \, {\rm nA}^2 / {\rm Hz}
    \frac{ I }{ {\rm A} }
    ,
\label{eq:shot-noise-level}
\end{equation}
where $e$ is the charge quantum. Note that currents
in a solid wire can have fluctuations below the shot noise limit
because the Coulomb interaction correlates the electrons. In
the following, we use $SN_{I}$ as a convenient
reference value.

If the wire current in a side guide
has a noise spectrum $S_{I}( \omega )$, it creates a magnetic field
with
\begin{equation}
    S_{B}( {\bf r}, \omega ) =
    \frac{ \mu_{0}^2 e I }{ 4 \pi^2 h^2 } \frac{ S_{I}( \omega ) }{ SN_I }
    .
\label{eq:technical-noise}
\end{equation}
This scalar quantity gives one component of the noise tensor.
The spin flip rate is then of the order of
\begin{equation}
    \Gamma_{\rm i \to f} \sim
    1\,{\rm s}^{-1}
    \frac{(\mu / \mu_{B})^2 }{ ( h / \mu{\rm m} )^2 }
    \frac{S_{I}( \omega_{L} ) }{ SN_{I} } \frac{ I }{ {\rm A} }
    .
\label{eq:estimate-tech-flip}
\end{equation}
We conclude that miniaturization beyond the micrometer scale
requires extremely low current noise to achieve trap lifetimes
longer than seconds. It may well turn out that very large scale
atom chip integration is only possible with static magnetic fields,
generated by magnetized nanostructures \cite{Hinds99b}.
Experimental trap lifetime data \cite{Reichel01b,Zimmermann02a}
show that currently used power supplies are quiet enough not to
reduce lifetimes at distances above $\sim10\,\mu$m.

\section{Spin coherence}
\label{s:spin-coherence}

Coherent manipulation on atom chips requires both that atoms
stay trapped and that their quantum state be preserved.
Here, we focus on the influence of magnetic noise on the
internal (spin) states. For example, different magnetic sublevels 
or hyperfine states are interesting candidates
to implement a bit of quantum information (qubit).
Noise-induced transitions between sublevels erase the qubit,
and this occurs on the same timescale as the spin flips
discussed in the previous section. But the information contained
in quantum superpositions can also be lost by pure `dephasing',
without changing sublevel populations \cite{Stern90}. 
For this process, which opens an additional channel for decoherence,
longitudinal magnetic fields, i.e. polarized along the static trapping field
(`phase noise'), come into play.

Consider two magnetic sublevels $|m_1\rangle$ and $|m_2\rangle$
that are simultaneously trapped.
A magnetic field fluctuation parallel to the
static trapping field changes the energy difference between
the two qubit states by
\begin{equation}
\Delta E( t ) = \Delta \mu_\Vert B_\Vert( {\bf r}, t )
,
\end{equation}
where $B_\Vert$ is the longitudinal magnetic field and
$\Delta \mu_\Vert = \langle m_2 | \mu_\Vert | m_2 \rangle
- \langle m_1 | \mu_\Vert | m_1 \rangle$ the differential
magnetic moment. (This difference can be substantially reduced
in alkalis by choosing hyperfine states that only differ
in the nuclear spin state.)
Let us now expose a superposition of $| m_1 \rangle$ and
$| m_2 \rangle$ to pure magnetic phase noise during an interaction
time $t$. The off-diagonal element of
the corresponding $2\times 2$ density matrix is then proportional
to \cite{Stern90}
\begin{equation}
\Big\langle
\exp \frac{ - {\rm i} }{ \hbar }
\int\limits_0^t\!{\rm d}t' \,
\Delta E( t' )
\Big\rangle
=
\exp\Big[
- \frac{ \Delta \mu_\Vert^2 t }{ 2 \hbar^2 }
S_\Vert( {\bf r}; 0 )
\Big]
,
\label{eq:qubit-dephasing}
\end{equation}
where $S_\Vert( {\bf r}; 0 )$ is the low-frequency limit of the
magnetic noise spectrum. (More precisely, one needs the spectrum
averaged over the frequency range $0 \ldots 1/t$. We neglect this
complication since the relevant spectra are flat in this range.)

From Eq.(\ref{eq:qubit-dephasing}), we conclude that
dephasing leads to exponential decoherence of
qubit superpositions with a rate
similar to the spin flip rate (\emph{cf.}\
Eq.(\ref{eq:rate-spectrum})). The decoherence rate gets
smaller when the
logical states have the same magnetic moment (reducing
$\Delta \mu_{\Vert}$), or when the magnetic field component
$B_{\Vert}$ shows much less noise.

Near field noise is rather
isotropic, as shown by the spectrum~(\ref{eq:nf-noise-spectrum}),
and therefore contributes equally to spin flips and phase noise.
Noise in the wire current gives only fluctuations perpendicular
to the guide axis so that dephasing is suppressed close to the
guide center. Note that this suppression is not complete because
of the finite, transverse width of the trapped wave function. 
In addition, gravity can displace
the actual trap center with respect to the magnetic field minimum
(the `gravitational sag' familiar from Bose condensates
in magnetic traps, see also \cite{Ketterle02a}).
In the harmonic approximation, a simple calculation leads to a 
reduction of the dephasing rate 
by a factor $(M g / \mu_\Vert b )^2$ where in order of magnitude
\begin{equation}
\frac{ M g }{ \mu_\Vert b }
\sim
0.1 \frac{ (M / {\rm amu})  [ g / (10\,{\rm m/s}^2) ] }{
(\mu_\Vert / \mu_{\rm B}) [ b / ({\rm G / cm}) ] }
.
\end{equation}
This ratio can be made quite small ($< 10^{-6}$) using typical
magnetic gradients $b = B_b / h$ achievable with the side guide.
Magnetic near fields thus remain as the main source of dephasing
noise, with a rate basically scaling like the spin flip rate.

\section{Vibrational coherence}

In this section, we turn to scattering processes that leave the atoms
in the magnetic trap, but perturb their center-of-mass motion.
This occurs whenever the magnetic noise is not spatially homogeneous.
A simple model is suggested showing
that the typical length scale for magnetic inhomogeneities
is of the order of the trap height $h$ for
both near field noise and current noise at the shot noise level.
We discuss the relation between random changes in the atoms' momentum
and the decoherence of their density matrix in position space.

\subsection{Scattering rate}

We again use Fermi's Golden Rule~(\ref{eq:Gamma-if-Fermi}), but now
the initial and final states are given by wave functions
$\psi_{\rm i,f}( {\bf x} )$. In the following, we retain only a single
trapped magnetic sublevel $|m_{\rm f}\rangle = |m_{\rm i}\rangle$
and assume that
the magnetic moment has the same orientation for all relevant
center-of-mass states. Its matrix
elements then reduce to $\mu_{\Vert}$
(the component along the trapping field). Writing out the
overlap integral between wave functions, the transition
rate~(\ref{eq:Gamma-if-Fermi}) becomes
\begin{equation}
    \Gamma_{\rm i\to f} =
    \frac{ \mu_{\Vert}^2 }{ \hbar^2 }
    \int\!{\rm d}^3x \, {\rm d}^3x' \,
    M_{\rm f\-i}^*( {\bf x} ) M_{\rm f\-i}( {\bf x}' )
    S_{\Vert}( {\bf x}, {\bf x}'; - \omega_{\rm f\-i} )
    ,
    \label{eq:gamma-overlap}
\end{equation}
where the wavefunction overlap is given by
\begin{equation}
M_{\rm f\-i}( {\bf x} ) = \psi_{\rm f}^{*}( {\bf x} ) \,
\psi_{\rm i}( {\bf x} )
,
\end{equation}
the transition energy $\hbar\omega_{\rm f\-i}$ is the difference
between the center of mass levels,
and the field correlation spectrum is the generalization
of the noise spectrum~(\ref{eq:def-noise-spectrum}):
\begin{equation}
S_{\Vert}( {\bf x}, {\bf x}'; \omega ) =
\int\limits_{-\infty}^{\infty}
\!{\rm d}t \,
{\rm e}^{ {\rm i}\omega t }
\sum_{\rm I}
p( {\rm I} )
\langle {\rm I} |
B_\Vert( {\bf x}, t )
B_\Vert( {\bf x}', 0 )
| {\rm I} \rangle
.
\label{eq:def-corr-spectrum}
\end{equation}
A useful figure that can be extracted from this
function is the correlation
length $l_{\rm c}$ that governs the variation
of~(\ref{eq:def-corr-spectrum}) as a function of
distance ${\bf s} = {\bf x} - {\bf x}'$. As a
general rule, one has $S_{\Vert}( {\bf x}, {\bf x}'; \omega ) \to
0$ if $s \gg l_{\rm c}$ because the fields
$B_{\Vert}( {\bf x}, t)$ and $B_{\Vert}( {\bf x}', t)$
become decorrelated. More quantitatively, we define
here $l_{\rm c}$ by the following expansion for small deviations
from the trap center ${\bf r}$ 
\begin{eqnarray}
&&
S_\Vert( {\bf x}, {\bf x}'; \omega ) \approx
S_\Vert( {\bf r}; \omega )
\Big[ 1 - \frac{ ({\bf x} - {\bf x}')^2 }{ l_{\rm c}^2 }
\Big]
\label{eq:def-lc}
\\
&&
\mbox{provided }
|{\bf x} - {\bf r}|, \, |{\bf x}' - {\bf r}| \ll l_{\rm c}
.
\nonumber
\end{eqnarray}
In Sec.\ref{s:noise-lc}, we show that the correlation length
$l_{\rm c}$ is comparable to the trap height $h$, for both
near field and current noise.

\subsection{Heating}

Consider the vibrational motion along one transverse direction
(the $x$-axis, say) in the harmonic region of the side guide. 
Magnetic noise can induce transitions
between different quantum states in the trap (`heating').
The transition $0 \to 1$ between the ground and
first excited states is particularly interesting and
involves the overlap integral
($\omega_{\rm f\-i} = \Omega$)
\begin{equation}
\int\!{\rm d}x \, {\rm d}x' \,
    M_{\rm f\-i}^*( x ) M_{\rm f\-i}( x' )
    S_{\Vert}( {\bf x}, {\bf x}'; - \Omega )
\approx
\frac{ a^2 }{ l_{\rm c}^2 }
S_{\Vert}( {\bf r} ; - \Omega )
\label{eq:Lamb-Dicke}
\end{equation}
where $a = (\hbar / 2 M \Omega )^{ 1/2}$ is the (rms) size
of the trap ground state ($M$ is the atomic mass)
and the noise spectrum is evaluated
at the trap center ${\bf r}$. Eq.(\ref{eq:Lamb-Dicke}) is derived
in the limit of strong confinement ($a$ much smaller
than the correlation length $l_{\rm c}$),
using the expansion~(\ref{eq:def-lc}).

Comparing~(\ref{eq:Lamb-Dicke}) to~(\ref{eq:rate-spectrum}),
we conclude that heating in tight traps is
suppressed relative to spin flips, due to
the small ratio $(a/l_{\rm c})^2 \ll 1$.
We stress that this is not due to the increase in
the trap frequency (because the noise is essentially white),
but due to the small size of the trap ground state.
Since $l_{\rm c} \sim h$, the excitation rate $\Gamma_{0\to 1}$
follows power laws $1/h^3$, $1/h^4$, instead of the flip
rates~(\ref{eq:estimate-gamma}, \ref{eq:estimate-tech-flip}),
depending on the noise source. For example, the
rate due to near field noise above a metallic half-space
scales like \cite{Folman02}
\begin{equation}
\Gamma_{0\to1} \simeq
\frac{ 1{\rm s}^{-1} \,( \mu / \mu_B )^2
(T_s / 300{\rm K}) }{ ( M / {\rm amu} )
( \Omega / 2\pi \, 100{\rm kHz} )
( \varrho / \varrho_{\rm Cu} )
( h / \mu{\rm m} )^3 }
.
\label{eq:estimate-substrate-heating-rate}
\end{equation}
Heating is also relevant for the decoherence of qubits
implemented in different vibrational levels. One can derive a master equation
for the density matrix in the harmonic well that shows that the corresponding
off-diagonal elements decay at a rate comparable to $\Gamma_{0 \to 1}$.
For details, see \cite{Henkel99c,Turchette00a}.

Finally, we note that even spatially homogeneous magnetic
fluctuations can induce heating when they change the trap
position or curvature. Displacing the trap is equivalent to a
force and drives the transition $0 \to 1$ with a rate
\cite{Folman02,Savard97,Gehm98}:
\begin{equation}
\Gamma_{0\to 1} = \frac{ M \Omega^3 }{ 2 \hbar }
S_{h}( - \Omega )
\end{equation}
where $M$ is the atomic mass and
$S_{h}( \omega )$ is the spectrum of the trap height
fluctuations. Taking into account only field fluctuations
caused by the wire current, we get in order of magnitude
\begin{eqnarray}
\Gamma_{0\to 1} &\sim&
3 \,{\rm s}^{-1}
(M / {\rm amu} ) ( \Omega / 2\pi \,100\,{\rm kHz} )^3
\nonumber
\\
&& \times
\frac{ I / {\rm A} }{ (B_b /  {\rm G} )^2  }
\frac{ S_I( \Omega ) }{ SN_I }
.
\label{eq:estimate-heating-current-noise}
\end{eqnarray}
This rate is still reasonably small for sufficiently large
bias fields $B_b > 50\,$G. Note, however, that a very strong
confinement may not be possible due to the increase
with the trap frequency $\Omega$: this occurs because
a larger spring constant $M\Omega^2$ translates position
fluctuations into larger forces.

Fluctuations in the trap position due to technical noise can
be reduced by correlating the currents in the wire with those
procuding the bias fields. The trap curvature, however, then still
fluctuates and changes the oscillation frequency $\Omega$.
This gives a parametric resonance on the $0 \to 2$ transition
with a rate \cite{Folman02,Savard97,Gehm98}:
\begin{eqnarray}
\Gamma_{0\to 2} &= &\frac12 S_{\Omega}( -2\Omega )
\\
&\sim &
3 \times 10^{-8}\,{\rm s}^{-1}
\frac{
( \Omega / 2\pi \,100\,{\rm kHz} )^2}{ I / {\rm A} }
\frac{ S_I( 2 \Omega ) }{ SN_I }
,
\end{eqnarray}
which is substantially smaller than the 
rate~(\ref{eq:estimate-heating-current-noise}).
Fluctuations in $\Omega$ also induce phase noise on the quantum
states in the harmonic trap because their energy difference
is $\hbar \Omega$.
Arguing as in Sec.\ref{s:spin-coherence},
one finds a dephasing rate comparable to $\Gamma_{0\to 2}$.

In experiments with trapped atoms, it is relatively simple to observe the rate of
temperature increase, $\dot T$. In the harmonic
approximation, one finds a `heating rate'
$k_B \dot T = \hbar\Omega \Gamma_{0\to 1}$
for noise driving the transitions $n \to n \pm 1$ \cite{Savard97,Gehm98}.
Currently observed values are in the range of $0.05 - 1\,\mu$K/s
\cite{Reichel01b,Zimmermann02a}.
This is orders of magnitude larger than predicted
by~(\ref{eq:estimate-substrate-heating-rate}) and excludes
an origin dominated by near field noise.
Heating can probably be attributed to current fluctuations,
as given by~(\ref{eq:estimate-heating-current-noise}), assuming
typical power supply noise spectra (above shot noise).
Ambient electromagnetic noise (`electrosmog') may also play a role 
and is currently under investigation.

\section{Spatial coherence}

\subsection{Scattering `cross section'}

We now consider the quasi-free motion along the side guide axis $Oz$,
the transverse motion is assumed to be `frozen out'.
Noise induces scattering $p_{\rm i} \to p_{\rm f}$
between different momentum states, where the
wavefunction overlap
$M_{\rm f\-i}( z ) = L^{-1}\,\exp( -{\rm i} q_{\rm f\-i} z )$
involves the wavevector transfer $\hbar q_{\rm f\-i} =
p_{\rm f} - p_{\rm i}$ ($L$ is a normalization length).
The transition frequency is $\omega_{\rm f\-i} = q_{\rm f\-i} p_{\rm i} / M
+ \hbar q_{\rm f\-i}^2 / 2 M$.
We end up with the spatial Fourier transform of the
correlation function (the vectors ${\bf x}$, ${\bf x}'$ only differ
in their $z$-components)
\begin{eqnarray}
\Gamma_{\rm i\to f} &=&
    \frac{ \mu_{\Vert}^2 }{ \hbar^2 }
    \int\!\frac{ {\rm d}z \, {\rm d}z' }{ L^2 } \,
    {\rm e}^{ {\rm i} q_{\rm f\-i} ( z - z' ) }
    S_{\Vert}( {\bf x}, {\bf x}'; - \omega_{\rm f\-i} )
    \nonumber\\
&\approx&
    \frac{ \mu_{\Vert}^2 }{ \hbar^2 }
    S_{\Vert}( {\bf r} ; - \omega_{\rm f\-i} )
    \int\!\frac{ {\rm d}s }{ L } \,
    {\rm e}^{ {\rm i} q_{\rm f\-i} s }
    C( s; - \omega_{\rm f\-i} )
.
\label{eq:prefactor-Gamma}
\end{eqnarray}
To perform the last step, we have assumed that the correlations
involve only the distance $s = z - z'$ (statistically homogeneous
noise) and used the normalized correlation function
($C( 0 ; \omega ) \equiv 1$):
\begin{equation}
C({z} - {z}'; \omega ) = \frac{ S_{\Vert}( {\bf x}, {\bf x}'; \omega ) }{
S_{\Vert}( {\bf r} ; \omega ) }
.
\end{equation}
Finally, we get a transition rate per wavevector transfer
${\rm d}\Gamma / {\rm d}q$ by dividing by the wavevector spacing
${\rm d}q = 2\pi/L$ in the quantization volume:
\begin{equation}
\frac{ {\rm d}\Gamma_{\rm i\to f} }{ {\rm d}q } =
    \gamma
    \int\!\frac{ {\rm d}s }{ 2\pi } \,
    {\rm e}^{ {\rm i} q_{\rm f\-i} s }
    C( s; - \omega_{\rm f\-i} )
,
\label{eq:cross-section}
\end{equation}
where the rate $\gamma$ is the prefactor of the integral
in Eq.(\ref{eq:prefactor-Gamma}). If the elements of the magnetic noise
tensor are of comparable magnitude, this rate is of the same order
as the spin flip rate~(\ref{eq:rate-spectrum}).
As mentioned at the end of Sec.\ref{s:spin-coherence},
this is the case for near field noise
and to a some extent also for fluctuations due to current noise.

We conclude from the differential scattering
rate~(\ref{eq:cross-section})
that typical momenta exchanged with the noise field have a magnitude
$q_{\rm f\-i} \sim \hbar / l_{\rm c}$
given by the inverse noise correlation length.

\subsection{Decoherence}

The relation to spatial decoherence of the atoms
has been made more quantitative in \cite{Henkel01a},
where the following master equation for the density matrix
$W(z, p )$ in the Wigner
representation is derived
\begin{eqnarray}
&&\Big(
\partial_t + \frac{ p }{ M }
\partial_z \Big)
W( z, p )
\nonumber
\\
&& =
\int\!{\rm d}q \,
\frac{ {\rm d}\Gamma }{ {\rm d}q }
\Big(
W( z, p + \hbar q )
-
W( z, p )
\Big)
.
\label{eq:master-equation}
\end{eqnarray}
The scattering integral on the right hand side describes processes
$p \leftrightarrow p + \hbar q$,
while the left hand side gives the free ballistic
motion along the guide axis.

The master equation~(\ref{eq:master-equation})
can be solved exactly using the fact that the
`scattering cross section'~(\ref{eq:cross-section}) is essentially
independent of the transition frequency $\omega_{\rm f\-i}$ for
the relevant magnetic field fluctuations. One finds the following
expression for the spatially averaged coherence function of the
atoms \cite{Henkel01a}
\begin{eqnarray}
\rho( s, t ) &\equiv& \int\!{\rm d}z
\langle \psi^*( z + s, t ) \psi( z, t ) \rangle
\label{eq:def-cohfunc}
\\
&=&
\rho( s, 0 ) \exp\Big( - \gamma t \, [ 1 - C( s; 0 )] \Big)
,
\label{eq:spatial-decoherence}
\end{eqnarray}
where the brackets denote the average with respect to the noise
and $C( s; 0 )$ is the low-frequency limit of the noise correlation
function.

From the coherence function~(\ref{eq:spatial-decoherence}),
we identify $\gamma [ 1 - C( s; 0 )]$ as the decoherence rate for
spatially separated superposition states:
for a splitting greater than the correlation length, $s \gg l_{\rm c}$,
the correlation function $C( s; 0)$ is zero and
the superposition decays into
a statistical mixture on a time scale given by $1/\gamma$,
comparable to the spin lifetime.
Superpositions with smaller splitting $s$ decay more slowly, with
a rate scaling like
$[ 1 - C(s; 0)] \gamma \approx (s/l_{\rm c})^2 \gamma \ll \gamma$,
using the expansion~(\ref{eq:def-lc}).
This behaviour was also found in a decoherence model
by W.\ H.\ Zurek \cite{Zurek91} who used a master equation in
Fokker-Planck form instead of our Eq.(\ref{eq:master-equation}).

We note that the decoherence rate just found also describes
the dephasing between the arms of a guided matter wave
interferometer \cite{Folman02,Hinds01a,Andersson02}.
This follows from an argument similar to
that used in Eq.(\ref{eq:qubit-dephasing}). In this context,
$s$ is the (transverse) separation between the arms of the
interferometer. For more details, we refer to \cite{Folman02}.

\subsection{Decoherence of a condensate}

We now discuss the extension of the previous results to the
case of a Bose condensate in a linear waveguide
(see \cite{Reichel02a} for a review of experiments). 
The single-particle wave functions then have to be replaced 
by collective modes, and the effective 
potential is changed due to atom-atom interactions. For example,
the (transverse) ground state has a larger width compared
to the single-particle wave function so that the
atomic spins are no longer aligned along the guide axis.
We present here preliminary results for the
decoherence of a Bose condensate focussing on a 
quasi-1D regime \cite{Olshanii98,Ketterle01b,Dettmer01,Petrov01}.
Adopting a mean-field description, we have
performed Monte-Carlo simulations of the one-dimensional
Gross-Pitaevski equation
\begin{equation}
{\rm i} \hbar \frac{ \partial \psi }{ \partial t }
=
- \frac{ \hbar^2 }{ 2 M }
\frac{ \partial^2 \psi }{ \partial z^2 }
+
V( z, t ) \psi
+
g |\psi( z ) |^2 \psi
\end{equation}
where the coupling constant $g = 2\hbar\Omega a_{\rm s}$
is proportional to
the (transverse) trap frequency and the s-wave scattering length
$a_{\rm s} > 0$
(repulsive interactions).
The random potential $V( z, t )$ is chosen in accordance with the
correlation functions relevant for atom chip traps (white noise and
lorentzian spatial correlations).
The initial situation is a condensate in the ground
state of a harmonic trap superimposed  on the wave\-guide potential.
Quantum (phase) fluctuations \cite{Dettmer01,Petrov01} are ignored
assuming effectively $T_{\rm at} = 0$.
The harmonic confinement is instantaneously released at
$t = 0$, and the cloud expands along the waveguide axis.

\begin{figure}[h]
\centerline{%
\resizebox{100mm}{!}{%
\includegraphics*{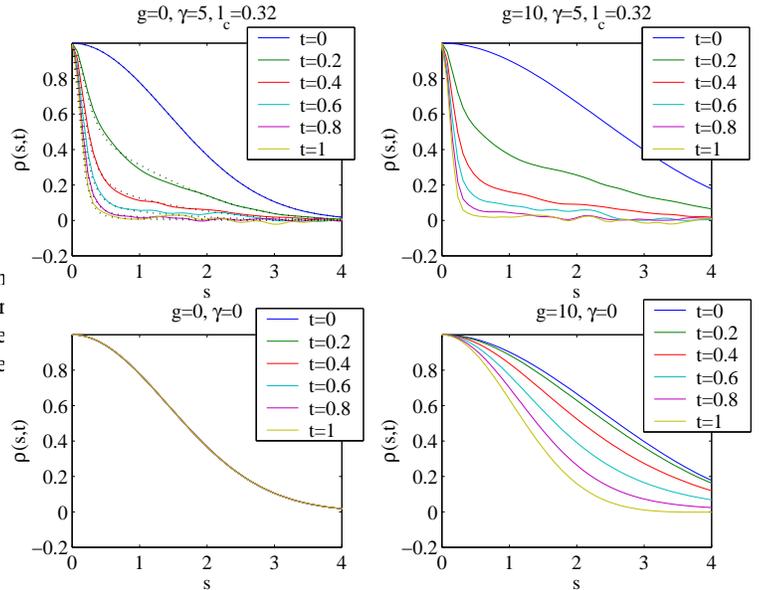}}}
\caption[]{Spatially averaged coherence function~(\ref{eq:def-cohfunc})
of a condensate expanding in a noisy waveguide.
Upper row: with noise giving a total scattering rate
 $\gamma = 10$ and correlation length $l_{\rm c} = 1/\sqrt{10}$.
Lower row: no noise ($\gamma = 0$). Left column:
noninteracting cloud ($g = 0$), right column: with interactions
($g = 10$). 
The dotted lines in the upper left panel are the
analytical prediction~(\ref{eq:spatial-decoherence}).
The units are harmonic oscillator units relative to the initial
confinement along the guide axis.
}
\label{fig:bec.eps}
\end{figure}

The simulation results given in figure~\ref{fig:bec.eps}
show the spatially averaged coherence function of the condensate,
Eq.(\ref{eq:def-cohfunc}),
as a function of the separation $s$ for different expansion times $t$.
One observes that in the presence of noise
(scattering rate $\gamma \ne 0$), the coherence length (the width
of $\rho(s, t)$) is reduced as time increases
--- the cloud breaks apart in mutually incoherent patches
that have `seen' different noise potentials.
The dotted lines in the upper left figure give the
prediction~(\ref{eq:spatial-decoherence}) of the master equation
for noninteracting atoms,
and we note a very good agreement between the analytics and the
numerical data.

We find that the decoherence scenario is not qualitatively
changed by a moderate self-interaction,
as a comparison of the panels for $g = 0$ and $g = 10$ shows.
Only for short times is the coherence length of an interacting
cloud larger because the ground state is broadened by the
interactions. A more detailed investigation
of condensate decoherence, including analytical approximations
and the limit of strong interactions, will be presented elsewhere
\cite{Henkel02c}.

\section{Noise correlation length}
\label{s:noise-lc}

We finally show that magnetic noise involved in atom chips has
a correlation length $l_{\rm c} \sim h$, for both near field
and current noise.

\subsection{Near field noise}

The spatial correlation function for the electric
near field above a thermal, planar substrate was studied in \cite{Henkel00b}.
A similar calculation yields for the magnetic near field the
following normalized correlation function:
\begin{equation}
C(s; 0) = \frac{ 8 h^2 }{
(2 h + \sqrt{s^2 + 4 h^2})
\sqrt{s^2 + 4 h^2} }
\approx
\frac{ 16 h^2 / 3 }{
s^2 + 16 h^2 / 3 }
\label{eq:nf-corr}
\end{equation}
This gives the correlation for the magnetic field component
$B_{\Vert}$, taken at positions $z$, $z'$ on the side guide
axis with a separation $s$.
We have assumed that the field wavelength
(fixed by the transition frequency $\omega_{\rm f\-i}$)
is much larger than the relevant distances $s, h$.
The lorentzian form in Eq.(\ref{eq:nf-corr})
is a good approximation for all distances
where $C(s; 0)$ is sensibly nonzero
and shows even more explicitly
that the correlation length is of the order of the guide height $h$.
This is because the noise fields radiated by each
volume element of the substrate are quasi-static in the near field
and decay algebraically (no retardation).

\subsection{Shot noise correlations}

One might think at first sight that
magnetic noise due to current fluctuations
should have a large correlation length because the relevant
electromagnetic frequencies (kHz to MHz range)
propagate with a large wavelength along the wire. The analysis
of near field fluctuations has shown, however, that the wavelength
is not really the relevant scale as soon as one is sensitive to
non-propagating fields generated by nearby sources.
For this reason, we suggest a simple toy model for the flow of electrons
through a thin wire that allows to recover both the noise spectrum
of the magnetic field and its spatial correlations.

The ingredients of the model are sketched in
figure~\ref{fig:wire.coords.eps} and details are given in
Appendix~\ref{a:shot-noise}.
The electrons are assumed to move independently and
ballistically (the Drude electron gas model),
their transverse position in
the wire is neglected compared to the guide distance $h$.
It is beyond the scope of this model to describe 
correlations between the electrons that
could lead to lower current fluctuations.
Neither does the model describe diffusive electron transport,
we comment on that below.

\begin{figure}[h]
\centerline{%
\resizebox{80mm}{!}{%
\includegraphics*{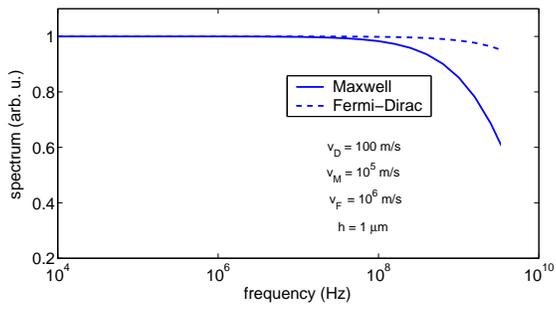}}}
\caption[]{Magnetic noise spectrum~(\ref{eq:shot-noise-result})
from the ballistic electron model.
The spectrum is taken at the side guide center (height $h = 1\,\mu{\rm m}$)
and normalized to its low-frequency limit.
The electron velocity distribution $P(v)$ is that of Maxwell
or Fermi-Dirac with characteristic velocities as given in the
figure. The distribution is centered around the drift velocity
$v_D$.}
\label{fig:shot.spec.eps}
\end{figure}

The result of the model is the
magnetic noise spectrum~(\ref{eq:shot-noise-result}) given
in the Appendix.
We recover the previous spectrum~(\ref{eq:technical-noise})
(with current noise at the shot noise level, $S_{I}( \omega )
= SN_{I}$) in the low-frequency
limit where $\omega h \ll v$ for all relevant electron velocities $v$.
At high frequencies, the noise is reduced because one needs
fast electrons to produce magnetic field `pulses' with short duration
$\sim h / v$.
This is visible in Figure~\ref{fig:shot.spec.eps}
where the spectrum~(\ref{eq:shot-noise-result}), for fixed
$h = 1\,\mu$m, is shown as a function of frequency,
normalized to its low-frequency limit.
The high-frequency cutoff occurs at $\omega h \gg v_{\rm F}$ where
$v_{\rm F}$ is the Fermi velocity when
the Fermi-Dirac distribution is taken for $P(v)$.
A Maxwellian distribution gives a similar behavior
with a lower cutoff, as shown in the figure.
Note again that at most frequencies relevant for atom chip traps,
the noise spectrum can be assumed flat.

The characteristic length scale $l_{\rm c}$ for the spatial correlation of the
shot noise fields can also be read off from
Eq.(\ref{eq:shot-noise-result}). In the directions perpendicular to the
guide axis, it is given by the guide height $h$,
and we recover the same correlation length as for near field
fluctuations. For the motion along the guide axis, the simplest way
is compute the differential
scattering rate~(\ref{eq:cross-section}).
For a process
$p_{\rm i} \to p_{\rm i} + \hbar q$, we get
\begin{equation}
\frac{
{\rm d}\Gamma_{\rm i \to f} }{ {\rm d}q } =
\gamma_{\rm SN}
\frac{ p_{\rm i} }{ M q }
P( p_{\rm i} / M )
\left[ q h K_1( q h ) \right]^2
,
\label{eq:result-shot-cross-section}
\end{equation}
where we have neglected the recoil shift $\hbar q^2 / 2M$ compared to
the Doppler shift $q p_{\rm i} / M$.
The scattering rate is
\begin{equation}
\gamma_{\rm SN} =
\frac{ \mu_\Vert^2 \cos^2 \alpha }{ \hbar^2 }
\frac{ \mu_0^2 e I }{ 4\pi^2 h^2 }
,
\label{eq:shot-noise-scattering-rate}
\end{equation}
where $\alpha$ is the angle between the atomic spin and the
wire field.
If the harmonic approximation for the transverse
motion is not valid, $\cos\alpha$ is not small 
and~(\ref{eq:shot-noise-scattering-rate})
is comparable to the spin flip rate (\emph{cf.}
Eq.\ref{eq:technical-noise}), typically a few $1\,{\rm s}^{-1}$.
$K_1$ in Eq.(\ref{eq:result-shot-cross-section})
is the modified Bessel function of the second kind,
and $P( v )$ is the electrons' velocity
distribution, taken at the atomic velocity $p_{\rm i} / M$.
It is interesting that the scattering
involves a class of electrons co-moving with the atom:
this suggests that the atomic wave in the linear guide
is diffracted by the spatially confined field pulse.
The divergence of the
cross-section~(\ref{eq:result-shot-cross-section})
for forward scattering ($ q \to 0$) is related to
the long range behavior of the field pulse;
this behavior also occurs for the Coulomb potential
where the Rutherford cross section diverges
in the forward direction.

We can now
conclude that also the longitudinal scattering is limited to
momentum transfers $\hbar q \le \hbar / h$.
This is due to the large-argument asymptotics of the Bessel function
in the result~(\ref{eq:result-shot-cross-section})
\begin{equation}
qh \gg 1: \quad
qh K_1( qh ) \approx \sqrt\frac{ \pi qh }{ 2 } {\rm e}^{ - qh }
,
\end{equation}
giving an exponential suppression for large $q h$. The shot noise field
is thus also `rough' on a scale $l_{\rm c} \sim h$, and does
not behave qualitatively different compared to the thermal
magnetic near field.

We finally comment on our neglect of the diffusive electron motion in the
wire. Since this effect gives rise to the nonzero resistivity of the wire,
it is in fact included in the Johnson noise approach for the metallic
substrate. The finite drift velocity of the electrons, which makes the
distribution $P(v)$ asymmetric, does not change our conclusions either
because it is typically much smaller than the width of $P(v)$.
The electron drift is taken into account
in the spectrum shown in figure~\ref{fig:shot.spec.eps}.

\section{Perspectives}

We have reviewed in this paper loss, heating and
decoherence mechanisms for wire-based atom chips
and their scaling with the microtrap
geometry and the substrate material properties. The importance of
the shot noise level for current noise has been highlighted.
Using a simple model, we have shown that the spatial correlation
length of magnetic fields due to shot noise is fixed by the
distance between microtrap and chip wire.

The extreme miniaturization of atom chip traps below the $1\,\mu$m
scale may not be possible with ``conventional", conducting
nanostructures, because magnetic field fluctuations due to thermal
and technical current fluctuations become quite strong. The key
process is a noise-induced change of the atomic sublevel
$|m\rangle$ (`spin flip'). Its rate, which can be related to the
trap lifetime, also dictates the order of magnitude of more subtle
processes involving heating of the center-of-mass motion or qubit
dephasing. The timescale for useful coherent manipulations
is thus at least limited to a few 100 ms at a height of a few $\mu$m.
Depending on the gate time of 2-qubit operations, this
time scale may be sufficient.

Several strategies leading to more robust atom chips can be imagined.
Cooled substrates reduce thermal near field noise, with a gain in lifetime
inversely proportional to the temperature. Quiet current drivers and/or
superconducting wires are an option to reduce current noise, possibly
below the shot noise level. Substrates with a permanent
magnetization may also provide the required low-noise environment.

The theory of magnetic noise close to complex, magnetized structures
can be developed starting from a description of the material in terms
of its electric and magnetic susceptibilities. According to the
fluctuation-dissipation theorem \cite{Varpula84,Agarwal75a},
the imaginary parts of these fix the magnitude of the noise current
and magnetization that generate thermal noise fields. This scheme
can be used at low frequencies (high temperatures), which is the
typical situation in atom chip traps, but can also be extended to
high frequencies where the noise reduces to vacuum fluctuations,
modified by the boundary conditions set by the substrate
(see e.g.\ \cite{Scheel00b} for a review). This kind of approach
can be used to compute atom-atom interactions mediated by
virtual photon exchange.

Another theoretical task is to estimate the coupling between
higher atomic levels and the electric and magnetic fields 
originating from the surface.
For example, it has been proposed to make
use of the electric dipole-dipole interaction to realize
controlled two-qubit gates on atom chip traps
\cite{Calarco00,Brennen00}. The gate operation is sped up when
the atoms are excited to high-lying Rydberg states, but these
states also interact strongly with the chip substrate due to
their large dipole moments, hence the need for a review of their
coherence times.

Finally, the understanding of the impact of the surface on the
fidelity of qubit operations needs to be studied in order to
optimize the construction of an atom chip quantum processor.

\bigskip\

\begin{small}
\noindent\textbf{Acknowledgments.}
C.H.\ took a great benefit from a discussion with Marc-Andr\'e
Dupertuis,
and thanks Sierk P\"{o}tting for his collaboration in earlier stages
and for providing figure~\ref{fig:guide.illu.eps}.
Thanks to Johannes Hecker Denschlag for being a critical co-author in
previous work.  The help of Simon A. Gardiner with the numerical
simulations of figure~\ref{fig:bec.eps} is gratefully acknowledged.
We also thank an anonymous referee for carefully reading the
manuscript. This work has been supported by the
Deutsche Forschungsgemeinschaft via the `Schwerpunktprogramm
Quanten-Informationsverarbeitung' and by the European Union
via the ACQUIRE network (contract number IST-1999-11055).
\end{small}

\appendix

\section{Correlation of magnetic fields generated by shot noise}
\label{a:shot-noise}

\begin{figure}[h]
\centerline{%
\resizebox{70mm}{!}{%
\includegraphics*{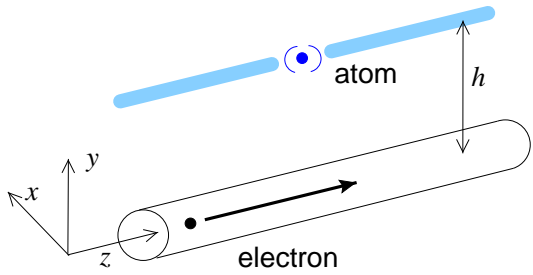}}}
\caption[]{Model for shot noise: ballistic electron flow.}
\label{fig:wire.coords.eps}
\end{figure}

Each electron (charge $e$), during its passage with velocity $v$
below the trapped atom, gives an electromagnetic `pulse'
whose vector potential is given by
\begin{eqnarray}
{\bf A}( {\bf x}, t ) &=&
{\bf e}_z \,
e \,k( {\bf x}, t - t_0; v )
\nonumber
\\
&=&
\frac{ {\bf e}_z \, \mu_0 e v / 4\pi }{
\left[
x^2 + y^2 + (z - v( t - t_0))^2 \right]^{1/2} }
,
\label{eq:A-pulse}
\end{eqnarray}
where the coordinates are chosen as shown in
figure~\ref{fig:wire.coords.eps}.
The atom actually experiences an average vector potential that is 
due to the flow of many electrons, passing below the atom
at random instants $t_0$. 
With the assumption of a stationary electron flow,
we find
\begin{equation}
\langle {\bf A}( {\bf x}, t ) \rangle =
- {\bf e}_z \frac{ \mu _0 }{ 4 \pi } I \,
\log( x^2 + y^2 ) + \mbox{const.}
\end{equation}
where the average current is $I = e \langle n \rangle / \Delta t$
with $n$ the number of electrons flowing during the interval $\Delta t$.
This average vector potential is time-independent and
gives the static magnetic field generated by the current.
Note that it does not involve the velocity distribution of
the electrons.
(The calculation is analogous to the calculation of a photodetector
current, as outlined in chapter 9.8 of \cite{MandelWolf}.)

We are interested in the correlation function of the vector potential
(from which the magnetic field correlations follow via differentiation).
Subtracting the average value and performing again the average over the
flowing electrons, we find using the approximation of independent electrons
\begin{eqnarray}
&&
\langle A_z( {\bf x}, t + \tau )
A_z( {\bf x}', t )
\rangle
-
\langle A_z( {\bf x}, t + \tau ) \rangle
\langle A_z( {\bf x}', t ) \rangle
\nonumber
\\
&& =
\frac{ e^2 \langle n \rangle }{ \Delta t }
\int\!{\rm d}t' {\rm d}v \, P( v )
k( {\bf x}, t' + \tau; v )
k( {\bf x}', t'; v )
,
\end{eqnarray}
where the `pulse function' $k( \ldots, t; \ldots )$ is defined in
Eq.(\ref{eq:A-pulse}).
The prefactor $e^2 \langle n \rangle / \Delta t = e I$ is
the shot noise spectrum $SN_I$ (Eq.\ref{eq:shot-noise-level}).
The Fourier integral with respect to the
time difference $\tau$ gives the squared Fourier transform of
the pulse function that can be evaluated analytically.
We do not give this formula here, but proceed directly to the magnetic
field correlation tensor. After differentiation, one needs the
Fourier transform
\begin{equation}
\int\!{\rm d}t
\frac{ v \, {\rm e}^{ {\rm i} \omega t }}{
\left[
r^2 + (z - v t )^2 \right]^{3/2} }
=
\frac{ 2\,{\rm e}^{ {\rm i} \omega z / v } 
|\omega/v| }{ r }
K_1( r |\omega/v| )
,
\end{equation}
where $r^2 = x^2 + y^2$ and $K_1$ is a Bessel function.
The magnetic noise field has the same orientation
(azimuthal) as the static field so that
in cylindrical coordinates, the correlation tensor
has a single nonzero element given by
\begin{eqnarray}
&& S_{\varphi\varphi'}( {\bf x}, {\bf x}'; \omega ) =
\frac{ \mu_0^2 \, SN_I }{ 4\pi^2 \, r r'}
\int\!{\rm d}v \, P( v ) \, {\rm e}^{ {\rm i} ( z - z' ) \omega / v }
\nonumber
\\
&& \quad {} \times
\frac{ r r' \omega^2 }{ v^2 }
K_1( r |\omega/v| )
K_1( r' |\omega/v| )
.
\label{eq:shot-noise-result}
\end{eqnarray}
In the low-frequency limit where $\omega h, \omega r \ll v$,
one gets $(r \omega / v ) K_1( r \omega/v ) \approx 1
+ {\cal O}( (r \omega / v )^2 )$. At the guide center
$r = r' = h, \, z = z'$, we then recover
the magnetic noise spectrum~(\ref{eq:technical-noise}) with
$S_I(\omega) = SN_I$.

The scattering cross section~(\ref{eq:cross-section}) involves
the Fourier transform of the correlation
function~(\ref{eq:shot-noise-result}) with respect to the
distance $s = z - z'$. This gives a $\delta$-function that permits
us to perform the integral over the electron velocities.
At $r = r' = h$, we get
\begin{eqnarray}
&&\int{\rm d}v \, P( v )
\,\delta\Big( \frac \omega v + q \Big)
\left[
\frac{ h \omega }{ v }
K_1( h |\omega/v| )
\right]^2
\\
&&= \frac{ |\omega| }{ q^2 }
P( -\omega / q )
[ q h K_1( q h ) ]^2
.
\end{eqnarray}
For the scattering process $p_{\rm i} \to p_{\rm i} + \hbar q$,
$\omega = - \omega_{\rm f\-i}$ is the negative kinetic energy
difference (see~Eq.\ref{eq:prefactor-Gamma}), so that
\begin{equation}
- \frac{ \omega }{ q } = \frac{ q p_{\rm i} + \hbar q^2 / 2 }{ M q }
\approx \frac{ p_{\rm i} }{ M }
,
\end{equation}
neglecting the recoil shift compared to the Doppler shift.
This approximation yields the scattering
probability~(\ref{eq:result-shot-cross-section}).

\clearpage

\end{document}